\documentstyle[12pt]{article}
\def\beq{\begin{equation}}   \def\eeq{\end{equation}}

\newcommand{\gsim}{\lower.7ex\hbox{$
\;\stackrel{\textstyle>}{\sim}\;$}}
\newcommand{\lsim}{\lower.7ex\hbox{$
\;\stackrel{\textstyle<}{\sim}\;$}}

\newcommand{\ra}{\rightarrow}

\newcommand{\La}{\overline{\Lambda}}

\newcommand{\as}{\alpha_s}

\newcommand{\GeV}{\,\mbox{GeV}}
\newcommand{\MeV}{\,\mbox{MeV}}

\newcommand{\matel}[3]{\langle #1|#2|#3\rangle}

\begin{document}

\def\lsim{\mathrel{\rlap{\lower3pt\hbox{\hskip0pt$\sim$}}
    \raise1pt\hbox{$<$}}}         
\def\gsim{\mathrel{\rlap{\lower4pt\hbox{\hskip1pt$\sim$}}
    \raise1pt\hbox{$>$}}}         

\begin{titlepage}
\renewcommand{\thefootnote}{\fnsymbol{footnote}}

\begin{flushright}
TPI-MINN-97/21-T\\ 
UND-HEP-97-BIG\hspace*{.2em}04\\
hep-ph/9706520\\
\end{flushright}
\vspace{.3cm}
\begin{center} \Large
{\bf The Hadronic Recoil Mass Spectrum in Semileptonic $B$ Decays and 
Extracting $|V_{ub}|$ in a 
Model-Insensitive Way}
\end{center}
\vspace*{.3cm}
\begin{center} {\Large
I. Bigi $^{a}$, R.D. Dikeman $^{b}$, N. Uraltsev $^{b,c}$\\
\vspace{.4cm}
{\normalsize
$^a${\it Dept.of Physics,
Univ. of Notre Dame du
Lac, Notre Dame, IN 46556, U.S.A.\\
$^b$ {\it  Theoretical Physics Institute, Univ. of Minnesota,
Minneapolis, MN 55455}\\
$^c$ {\it St.Petersburg Nuclear Physics Institute,
Gatchina, St.Petersburg 188350, Russia\footnote{Permanent address}
}}\\
}}
\vspace*{1.7cm}
{\small \it  Reported at {\rm 3rd} BaBar Physics Workshop, 
Orsay 16--19 June 1997}
\vspace*{1.3cm}

{\Large{\bf Abstract}}\\
\end{center}

We present an extended discussion of the previously noted possibility to
extract $|V_{ub}|$ from an analysis of the hadronic recoil mass
spectrum in $B\ra X_u\,\ell \nu$ decays. Invariant mass spectra
containing perturbative as well as nonperturbative  corrections are
given; their shape is manifestly sensitive to the three basic 
quantities $\mu _\pi ^2$, $m_b$ and $\alpha_s$, whereas  the total
integrated rate is much less so.  Only a small fraction of $b\ra u$
transitions generates a recoil  mass $M_X$ of at least $M_D$. Moreover
we find that the fraction of events with $M_X \leq 1.5\GeV$ (to
reject leakage  from $b \ra c$ due to measurement errors) exhibits
fairly little dependence on  $\mu_\pi^2$, $m_b$ and $\alpha_s$;   
$|V_{ub}|$ can then be extracted in a  largely model-insensitive way.  
This conclusion is based on the  applicability of the OPE 
to actual semileptonic $B$ decays. A direct cross-check of
this assumption and a determination of the required basic parameters  
of the heavy quark expansion will be possible in the future with more
experimental data.

\vspace*{.2cm}
\vfill
\noindent
\end{titlepage}
\addtocounter{footnote}{-1}

\newpage

\section{Introduction}

The KM parameters are fundamental quantities in the Standard
Model. They have to be determined as reliably as possible for
two reasons, one of
a theoretical and one of a more phenomenological nature:
(1) It
is hoped that a future more complete theory will enable us to
calculate these parameters.
(2) CP asymmetries observable in strange and
beauty decays are predicted in terms of these parameters.

{\em Numerical} precision in the extraction of these quantities is
highly desirable. Theoretical schemes involving quite
different dynamical scenarios at very high energies tend to yield
KM parameters that do not differ very much when probed at or
below the electroweak scale. Also, since some of the
predictions for CP asymmetries in $B$ decays can be made with high
{\em parametric} accuracy, one wants to translate this achievement
into high {\em numerical} precision. Such considerations suggest a
benchmark of better than $10\%$ in accuracy for $|V_{cb}|$ and
$|V_{ub}|$ as a desirable goal.

With the emergence and increasing sophistication of heavy quark
expansions (for a detailed evaluation, see \cite{rev})
one has been able
to translate ever more precise measurements into extractions of
$|V_{cb}|$ with a theoretical uncertainty of about $5\%$ and improving
-- something that could not have been expected a few years ago.

The situation is much less satisfactory for $|V_{ub}|$. We know
certainly that $|V_{ub}| \neq 0$ holds since (i) the decays
$B \ra \pi \ell \nu$, $\rho \ell \nu$ have been identified and
(ii) $B \ra X\,\ell \nu $ has been observed with lepton energies $E_\ell$
that
are accessible only if $X$ does {\em not} contain a charm hadron:
$E_\ell \geq (M_B^2 - M_D^2)/2M_B = 2.31 \GeV$.
However to translate these findings into reliable numbers 
concerning
$|V_{ub}|$
is a much more difficult task theoretically: On the one hand
the exclusive decays $B \ra \pi \ell \nu$, $\rho \ell \nu$ depend on
bound-state effects in an essential way (heavy quark symmetry
can be relied upon here to a considerably lesser degree
than in $B \ra D^{(*)}\ell \nu$ and  
although quark models have been employed, 
there is no reliable way for gauging their theoretical
uncertainties).
On the other hand in analyzing the endpoint spectrum for
charged leptons in the inclusive decays $B \ra X\, \ell \nu$ one
encounters
different sorts of systematic problems. Only a fairly small
fraction
of the charmless semileptonic decays $B \ra X_u \ell \nu$, namely
around $10\%$
or so, produce a charged lepton with an energy {\em beyond} that
possible
for $B \ra X_{c} \ell \nu$; in addition, the $b\ra c$ rate is so
much bigger than that for $b\ra u$ that leakage from it due to
measurement errors becomes a serious background problem;
furthermore the endpoint region is particularly sensitive to
nonperturbative dynamics.

The importance of one such effect, namely the motion of the decaying
heavy quark inside the hadron, was recognized  a long time ago
and
the concept of `Fermi motion' was introduced  into phenomenological
models, albeit in an ad-hoc fashion  \cite{AP,ACCMM}. It was later
pointed out that Fermi motion emerges naturally in a  dynamical
treatment that is genuinely based on QCD
implementing  the heavy quark expansion through an operator product
expansion  (OPE) \cite{prl}. Conceptually it is similar to leading 
twist effects in deep inelastic scattering (DIS). Yet
some subtle, though significant peculiarities arise making it
somewhat
different from the simple-minded treatment in quark  models. In
particular, it is not permissible to identify the basic
expectation
value $\mu_\pi^2$ (to be defined below) with what is
usually called the average Fermi momentum $p_F^2$, 
extract it from a fit to
$b\ra c$
transitions in the most popular AC$^2$M$^2$ model, and then apply it
at face value to $b\ra u$ decays. For $p_F^2$ is in general 
different from $\mu_\pi^2$ even in the context of the
AC$^2$M$^2$ model itself; furthermore QCD dictates using a somewhat
different
set of constraints than were historically used in the AC$^2$M$^2$ 
model
\cite{roman}.

Nevertheless, we
have learned more than just negative lessons. We know  how to express
the
total width  $\Gamma (B \ra X_{u}\ell \nu)$ reliably in terms of
$|V_{ub}|$ and we have found descriptions of the Fermi motion
that are -- while not unique -- at least fully consistent with
everything we know about QCD. The theoretical tools involved
have been described before \cite{tools}. Here we want to
concentrate
on how they can be applied in a practical way. We will
briefly discuss $\Gamma (B \ra X_{u}\ell \nu)$ before
analyzing in detail how to extract $|V_{ub}|$ from the hadronic
recoil mass spectrum in semileptonic $B$ decays.

\section{Total Semileptonic Widths}

The semileptonic widths $\Gamma (B\ra X\,\ell \nu)$ have been
calculated through order $1/m_b^3$ \cite{buv,prl,bds}; the
effect of the
cubic terms is particularly small, and we neglect it in the rest of
our
discussion.
The leading nonperturbative corrections are
expressed through expectation values $\mu_G^2$ and $\mu_\pi^2$
of the chromomagnetic and kinetic heavy quark operators, respectively 
\beq
\mu_G^2 \equiv \matel{B}{\bar b\, \frac{i}{2}\sigma G\,
b}{B}/2M_B\;,
\qquad\qquad
\mu_{\pi}^2\equiv \frac{1}{2M_B}
\matel{B}{\bar b\, (i\vec D)^2 \,b}{B}\;,
\eeq
with $\sigma G = \sigma _{\mu \nu}G_{\mu \nu}$, 
where $G_{\alpha \beta}$ is the gluon field strength tensor; 
$\vec D$ denotes the covariant derivative.
The measurement of $\Gamma_{\rm sl}(B)$ has allowed the most
reliable extraction of $|V_{cb}|$ \cite{rev}:
$$
|V_{cb}| = 0.0419 \cdot \left( \frac{{\rm BR}(B \ra X_c \ell \nu)}
{0.105}\right) ^{\frac{1}{2}}\cdot \left( \frac{1.55\, {\rm ps}}
{\tau _B}\right) ^{\frac{1}{2}} \times
$$
\beq
\left( 1 - 0.012 \cdot \frac{\mu _{\pi}^2 - 0.5\, {\rm
GeV}^2}
{0.1\, {\rm GeV}^2} \right) \cdot
(1 \pm 0.015_{|_{pert}} \pm 0.01_{|_{m_b}} \pm 0.012)
\label{VCB}
\eeq
The largest theoretical uncertainty resides in the value of
$\mu_{\pi}^2$ as shown explicitly. For the total width depends
mainly
on the difference $m_b - m_c$ the error of which  is controlled by
$\mu_{\pi}^2$; the  remaining uncertainty in $m_b$ is stated in
Eq.(\ref{VCB}). The fourth term in  the last bracket represents an
estimate for the unknown effects of order $1/m_Q^3$ (and
higher)
and of possible deviations from  quark-hadron duality. The
perturbative
error reflects the  uncertainty in the value of $\as$ and the
weight of
the higher order corrections.

$\Gamma (B \ra l \nu X_{u})$ can
be treated in complete analogy, and one finds \cite{upset,rev}:
\beq
|V_{ub}| = 0.00465 \cdot \left( \frac{{\rm BR}(B \ra X_u \ell \nu)}
{0.002}\right) ^{\frac{1}{2}}\cdot \left( \frac{1.55\, {\rm ps}}
{\tau _B}\right) ^{\frac{1}{2}} \cdot (1 \pm 0.025_{|_{pert}}
\pm 0.03_{|_{m_b}} )
\label{VUBTOTAL}
\eeq
The dependence on $\mu _{\pi}^2$ has practically dropped out;
the uncertainty introduced by $m_b$ is
larger than in the $b\ra c$ case, but still quite small.

The real and highly nontrivial challenge is then of an experimental
nature, namely how to measure $\Gamma (B \ra X_{u} \ell \nu)$.
The only feasible way would presumably be to find a kinematical discriminator 
between $b\ra u$ and $b\ra c$ transitions. This will be discussed 
next. Eq.~(\ref{VUBTOTAL}) shows
that the main uncertainty will be experimental rather than
theoretical.

\section{The Hadronic Recoil Mass Spectrum}

As stated in the Introduction, phenomenological models
include the motion of the heavy quark through a given  ansatz.
It was also recognized \cite{BARGER} that a
measurement of the
hadronic recoil spectrum in semileptonic $B$ decays
\beq
\frac{d}{dM_{X}} \Gamma (B \ra X\, \ell \nu )
\eeq
would offer intrinsic advantages for disentangling the $b\ra u$
transitions
relative to the conventional analysis of the charged lepton
spectrum $d\Gamma /dE_\ell(B\ra X \ell \nu)$ --
if it could be done experimentally. In the lepton spectrum one
can separate $b\ra c$ and $b\ra u$ kinematically only in the
small slice $2.31\GeV \leq E_\ell \leq M_B/2 \simeq 2.64\GeV$,
and not surprisingly $\sim 90\%$ of $b\ra u$ is hidden
underneath the dominant $b\ra c$ transition.
In $M_{X}$, on the other hand, kinematical separation can
be achieved in
the much larger range $M_{\pi} \leq M_{X} \leq M_D$.
One might expect -- and simple quark model computations
like those of \cite{BARGER}
bear out the fact -- that the bulk of the $b\ra u$ contribution lies
below
that for $b \ra c$ when expressed in terms of $M_{X}$. Yet  a more
sophisticated analysis is required to see to which degree this
holds
true. The phenomenological descriptions involve some  ad hoc
assumptions of uncertain numerical reliability, and they  suffer
from
obvious failures or at least limitations, for example  in their
description of the hadronic recoil spectrum for  $b \ra c\;$: the
observed
mass spectrum dominated by the two narrow $D$ and $D^*$ peaks does
{\em
not} emerge from the quark model descriptions -- instead these models
 yield 
very smooth functions for $b\ra u$ and $b\ra c$ channels alike. Yet the
QCD-based
treatment yields a different expansion in the two cases
\cite{motion}.
The best tool for
these studies is provided by the heavy quark expansion; first we
will
sketch this theoretical technology  as it applies here and then
analyze
$b\ra u$ transitions in detail.

\subsection{The Methodology}

The theoretical treatment of inclusive semileptonic decays of $B$
mesons
carries a distinct similarity to deep inelastic lepton-nucleon 
scattering -- this analogy can be pursued at great length.
One defines a hadronic tensor as the $B$ meson expectation value
for the transition operator \cite{KOY}
\beq
h_{\mu \nu}(q)= \frac{1}{2M_B} \matel{B}{\hat T_{\mu \nu}(q)}{B}
\eeq
\beq
\hat T_{\mu \nu}(q) = i \int d^4x {\rm e}^{\,-iqx}
T\{ J^{\dagger}_{\mu}(x) \, J_{\nu}(0)\} \;  , \;\; \;
J_{\mu} = \bar q \gamma _{\mu}(1-\gamma _5)b\;.
\label{TMUNU}
\eeq
The hadronic tensor $h_{\mu \nu}$ can be decomposed into five
different Lorentz covariants
$$ 
h_{\mu \nu} (q)= - h_1 g_{\mu\nu}(q_0,\,q^2) + h_2(q_0,\,q^2) v_\mu
v_\nu
- ih_3 (q_0,\,q^2) \epsilon_{\mu\nu\alpha\beta}v_\alpha q_\beta + 
$$
\beq 
+ h_4(q_0,\,q^2) q_\mu q_\nu + h_5(q_0,\,q^2) (q_\mu v_\nu + q_\nu
v_\mu)
\eeq
($v$ is the $4$-velocity of the decaying $B$),
from which one obtains structure functions in the usual
way:
\beq
w_i (q_0,\,q^2) = 2 \,{\rm Im}\,h_i(q_0,\,q^2)\;.
\eeq
All inclusive observables can be expressed in terms of these
$w_i$. For $l=e$ or $\mu$ with $m_l \simeq 0$ only $w_1$, $w_2$
and $w_3$ are actually relevant.

Applying the OPE to the product of currents in Eq.~(\ref{TMUNU})
one can express the structure functions through an infinite series
of expectation values of operators of higher and higher dimension.
The
coefficients become singular when one approaches  free-quark
kinematics. Therefore, even the limit $m_b\ra \infty$ does not allow
one
to evaluate the structure functions completely -- it would require
the
resummation of infinite series of equally important nonperturbative
contributions.

Nevertheless, with the large mass of the $b$ quark one can pick up 
the leading operators for a given type of singularity, a procedure
similar to resummation of the leading-twist contribution in DIS.
Their
effect is combined into the heavy quark distribution function which
replaces the Fermi motion wavefunction of the quark models.

For $b\ra c$ transitions (i.e. for $q=c$ in Eq.~(\ref{TMUNU})) we
have
one more tool at our disposal: heavy quark symmetry imposes certain
constraints on the properties of the structure functions.
It can be revealed directly in the QCD $1/m_Q$ expansion
through studying
the small velocity limit to derive SV sum rules \cite{optical},
without
an a priori appeal to the underlying picture of strong interactions.
Thus the
tools are prepared to calculate  (among other things)   the hadronic 
recoil mass spectrum for $b\ra c$ in a way that is fully  consistent with
QCD
and in agreement with the data -- in contrast to  what
phenomenological
models yield. One should keep in mind, though, that even the full
power of
the heavy quark symmetry still leaves large room for variations when
the effects of higher order in $1/m_c$ and/or the 
recoil velocity, are addressed.

With respect to $b\ra u$ the situation is a priori less favorable
since heavy quark symmetry cannot be applied to the final state in
$b\ra u$ and a small velocity treatment would make no sense!
Nonetheless, in terms of the ``global'' characteristics relevant for
inclusive decays, the heavy quark expansion yields many constraints.

Similar to DIS, the moments of the structure functions, i.e. their
integrals
with integer
powers of $q_0$ at fixed $q^2$, are given in terms of the
expectation
values of local heavy quark operators of increasing dimension.
De-convoluting these  moments one can  determine a heavy
quark distribution function $F(x)$ which encodes the effects of the
Fermi motion. The dimensionless parameter $x$  plays the role of the
primordial momentum of the $b$ quark normalized to
$\bar \Lambda = M_B - m_b$; x varies between $-\infty$ and  $1$. All
structure functions in the leading approximation can then be
expressed
in terms of $F(x)$:
\beq
w_i(q_0,q^2) \stackrel{{\rm leading \; twist}}{=} 
\int_{-\infty}^1 dx\;
w_i^{pert}\left(q_0-\frac{x\La}{M_B} \sqrt{q_0^2-q^2},\,q^2 \right)
\:
F(x)\left(1-\frac{x\La}{M_B}\frac{q_0}{\sqrt{q_0^2-q^2}} \right)
\,dx
\,.
\label{STRUCTFUNC}
\eeq
Here $w_i^{pert}$ is a parton structure function including 
perturbative corrections. The exact form of this relation is not
unique
as long as only the leading-twist effects are summed up.
Nevertheless,
this particular form has some advantages, one of them being its
transparent physical meaning 
(although it might not be quite obvious at first
glance). The limitations -- both of
theoretical and
practical nature -- due to discarding the subleading-twist
contributions and neglecting the actual dependence of $F(x)$ on the
velocity of the final-state hadronic system, are discussed 
in \cite{bsg,dist}.

As stated above, the moments $a_i$ of $F(x)$ are given by the
expectation values of local heavy quark operators. In practice
we know only the size of the first few moments; in the adopted
normalization one finds
\beq
a_0 = 1\; , \; \; a_1 = 0 \; , \; \;
a_2 = \frac{\mu _{\pi}^2}{3\bar \Lambda ^2}\;\; .
\label{MOMENTS}
\eeq
One then chooses a specific functional form for $F(x)$ and adjust
its
parameters to reproduce the phenomenologically deduced moments.
This procedure was performed in Refs.~\cite{bsg,dist} where the
following
ansatz was employed:
\beq
F(x) = \theta (1-x) {\rm e}^{\,cx}(1-x)^{\alpha}[ a + b(1-x)^k]\;\;.
\label{ANSATZ}
\eeq
With abundant experimental information one can in principle measure $F(x)$, for
example, in the inclusive decays $b\ra s+\gamma$.

Knowledge of the structure functions allows one to calculate all
inclusive
distributions. For example, the invariant mass squared of the
hadronic final
state has the following kinematic form:
\beq
M_X^2 = M_B^2 + q^2 - 2M_Bq_0 = [m_b^2 + q^2 - 2 m_bq_0]
+ 2(m_b - q_0)\bar \Lambda + \bar \Lambda ^2
\label{Mx2}
\eeq
Note the explicit term proportional to $\La$; it manifests the
$1/m_b$
nonperturbative effect in the average value of $M_X^2$ \cite{WA}.

Eq.~(\ref{STRUCTFUNC})
has a very intuitive quark model interpretation \cite{dist}.
Yet there are some
relevant subtleties  that have to be kept in mind for proper
understanding as explained in detail in \cite{optical,bsg}:

\begin{itemize}
\item
The nature of $F(x)$ changes even {\em qualitatively} when going
from
heavy quark masses, $m_c^2 \gsim \bar \Lambda m_b$ --  to light
ones --
$m_u^2 \lsim \bar \Lambda ^2$. One of the advantages of
the representation
(\ref{STRUCTFUNC}) is that the relations stated in
Eq.~(\ref{MOMENTS})
for the first three moments hold for an arbitrary quark mass in the
final state;
the high moments essentially depend on it (as well as on
$q^2$,
through the quark velocity).

\item
In the presence of gluon bremsstrahlung the hadronic recoil mass
$M_X$ for
$b\ra u$ can become much larger than $\sqrt{\bar \Lambda m_b}$
which would hold in the simplest quark picture.

\item
We have adopted Wilson's prescription for the OPE
where an energy scale
$\mu$ is introduced to separate long- and short-distance dynamics;
the former are lumped into the matrix elements of the local
operators while the latter are incorporated into their coefficients.
No observable can depend on $\mu$; yet as a practical matter it
has to be chosen such that perturbative as well as nonperturbative
corrections can be brought under theoretical control.
More specifically, the moments $a_i$, the dimensionful parameter
$\La$
as well as
the distribution function $F(x)$ itself
depend on $\mu$. Some care has to be applied in
keeping track of the $\mu$ dependence.

\end{itemize}
While the second point is completely obvious, the first one is not
and is actually quite mysterious from
the perspective of a quark model description.

Putting everything together we arrive at
$$
\frac{d\Gamma}{dM_X^2} = \frac{1}{2M_B}
\int dq^2 \int dx F(x) \left[ 1- \frac{x\bar \Lambda}{M_B}
\left(1 - \frac{4q^2M_B^2}{(M_B^2 + q^2 - M_X^2)^2}\right)
^{-1/2}\right] \times
$$
\beq
\times
\frac{d^2\Gamma^{\rm pert}}{dq_0\,dq^2}\left(
\frac{M_B^2+q^2-M_X^2}{2M_B}
-\frac{x \La}{M_B}\sqrt{ \left(
\frac{M_B^2+q^2-M_X^2}{2M_B}\right)^2  -q^2}
\,,\; q^2\right)\;.
\label{13}
\eeq

A few more notes are in order on how the
perturbative corrections are treated. We follow here
Ref.~\cite{dist} and
include the exponentiated one-loop corrections at a given $q^2$, 
evaluated with a fixed
rather than
a running $\alpha_s$, without an explicit infrared cutoff \cite{greub}. The
reasons behind
this approach are given in \cite{dist}. To be consistent, we
then employ the $b$ quark pole mass computed to one-loop accuracy with the
same (frozen)
coupling:
\beq
\tilde m_b \simeq m_b(\mu ) +
\frac{16}{9}\frac{\alpha_s}{\pi} \mu
\eeq
and likewise for the kinetic term
\beq
\tilde \mu _{\pi}^2 \simeq \mu _{\pi}^2(\mu ) -
\frac{4}{3}\frac{\alpha_s}{\pi} \mu ^2\;.
\eeq
In principle, the complete perturbative coefficient functions and the
distribution function separately do not allow considering the limit
$\mu\ra 0$ whereas only the convolution (\ref{13}) is $\mu$-independent.
Within the accuracy of our calculations, however, it does not pose
problems. In particular, discarding the running of $\as$ allows one to
perform integration over gluon momenta down to zero. In respect to the
moments of $F(x)$ we considered above -- as pointed out
in Ref.~\cite{optical} -- this 
corresponds to subtracting the would-be perturbative contribution to the
expectation values of local operators from the domain below $\mu$. This
contribution must be evaluated in the theory with  
frozen coupling. 
This subtraction is performed in the equations above. 

The approximations we made in the treatment of the perturbative corrections
are not expected to produce significant distortion. 
Even in $b\ra s +\gamma$ decays
for the actual $b$ quark mass 
the major impact was found to be due to the soft primordial
distribution \cite{bsg}. Moreover, even including the effects of 
$\as$ running the resulting photon spectrum (which measures the
recoil mass spectrum) changed only a little upon variation of $\mu$ and
upon incorporating the running of $\as$ -- while short-distance and 
long-distance parts separately changed radically. The stability in the
case of semileptonic decays must be even better, since due to the sizable
average value of $q^2$, the effective energy release is much smaller and
logarithms of the ratios of the momentum scales encountered in the
perturbative calculations become rather insignificant.

To summarize our brief review of the relevant methodology:
\begin{itemize}
\item
The observable distributions are described in terms of structure
functions which in turn are expressed -- to leading approximation --
through the convolution of a short distance rate and the heavy quark
distribution function $F(x)$.
\item
$F(x)$ encodes the main impact
of nonperturbative dynamics conventionally referred to as
Fermi motion. In principle, it can be reconstructed from its
moments. In practice,
only the first few moments are known. One then chooses some
reasonable ansatz for $F(x)$; its parameters are adjusted so as to
reproduce its known moments. While one cannot claim that this
specific form for $F(x)$ is derived from QCD in a {\em unique} manner,
it represents a dynamical and self-consistent realization of QCD
and its known constraints. Once more constraints become available,
they can in turn be implemented; i.e., the ansatz can be refined
step by step.
\item
While many results or expectations previously inferred from
phenomenological models re-emerge, it would be quite
inappropriate to say they were `reproduced'. Deriving them from
QCD proper represents significant conceptual progress; it has
also warned us against various potential pitfalls and sharpened
our vision regarding various subtleties that are quite significant
even quantitatively, yet had escaped notice before.
\item
The {\em shape} of the $b\ra u$ distributions in practical
implementation of this program is controlled by three
quantities, namely
\begin{itemize}
\item
$m_b$, the $b$ quark mass ;
\item
$\mu _{\pi}^2$,
the kinetic operator for the $b$ quark;
\item
the strong coupling, $\alpha_s$.
\end{itemize}
\end{itemize}

\subsection{Analysis of $b \ra u$ Recoil Mass Spectrum}

The tools appearing above are now applied. We use as
central values for the basic parameters
\beq
\tilde m_b = 4.82\GeV\,, \qquad \tilde \mu _{\pi}^2 = 0.4\GeV^2\,,
\qquad
\alpha_s = 0.3
\eeq
corresponding to the parameters of $F(x)$
\beq
[\alpha, \, b/a, \, c] = [ 0.5, \, 0.39, \, 1.75]
\eeq
in Eq.~(\ref{ANSATZ}); $k=1$ has been preset.
The resulting shape of the hadronic recoil spectrum is shown in the
solid curve in  Fig.~1. Adopting $\alpha_s = 0.1$ and $\as=0.5$ instead 
produces the dashed and dotted curves which do not differ radically.
The
fact that $d\Gamma/dM_X$ shows such low sensitivity to the value of
$\alpha_s$ provides us with an a posteriori justification for our
usage of a fixed coupling. This fact was expected. The main
effect of the running of $\as$ can come from the low-scale part of the
bremsstrahlung. The exponentiation of the soft corrections, however,
suppresses the domain where they are most pronounced.

With fixed $\as$ the effect of the exponentiation of the soft
contributions is not very strong. For example, in the purely
perturbative calculation of the hadronic mass distributions it shifts
the value $M_X$, below which $60\%$ of decay events are expected, from
$0.55\GeV$ down to $0.4\GeV$. This effect of softening the
$M_X$-distribution must be more prominent if one
uses larger or running $\as$.

As anticipated, the mass spectrum is very broad and extends even
beyond $M_D$ -- yet only a small fraction does so, namely
$\sim 10\% $. Due to measurement errors there will be a tail from
$b\ra
c$ transitions {\em below} $M_D$. To avoid this leakage one  can
concentrate on recoil masses below a certain value  $M_{\rm max} <
M_D$.
The actual choice of $M_{\rm max}$ is driven by   competing
considerations: the lower $M_{\rm max}$, the less leakage from
$b\ra c$
will occur -- yet the smaller the relevant statistics, as  expressed
through $\Phi(M_{\rm max})$, the fraction of $b\ra u$ events with
$M_X$
below $M_{\rm max}$:
\beq
\Phi(M_{\rm max}) = \frac{1}{\Gamma (b\ra u)}
\int _0^{M_{\rm max}} dM_X \frac{d\Gamma}{dM_X}
\label{FRACTION}
\eeq

There is a third consideration to be elaborated now. Since the
recoil spectrum is shaped by nonperturbative dynamics, it is
sensitive to the values of $\tilde m_b$ and $\tilde \mu _{\pi}^2$.
Varying $\tilde \mu _{\pi}^2$ between
$0.2$ and $0.6\GeV^2$ while keeping $\alpha_s = 0.3$ and
$\tilde m_b = 4.82\GeV$ fixed one obtains the three curves of
Figs.~3 and 4. Changing $\tilde m_b$ between $4.77$ and $4.87\GeV$, one
obtains
a somewhat larger difference in the spectrum -- as shown in Figs.~5 and
6.
With abundant statistics one can, in principle,  distinguish the
different curves and try to
extract $m_b$ and $\mu _{\pi}^2$ -- in addition to $|V_{ub}|$. Yet
it will take some time before such statistics become available.
Realistically, we expect here only a cross-check of the consistency
with other, dedicated experimental evaluations of these parameters.

In the
meantime one can quite profitably concentrate on analyzing
$\Phi(M_{\rm max})$, with the primary goal placed on determining 
$|V_{ub}|$. Since the theoretical uncertainty in the total width
$\Gamma_{\rm sl}(b\ra u)$ or in the ratio
$\Gamma_{\rm sl}(b\ra u)/\Gamma_{\rm sl}(b\ra c)$ is small and
depends on different type of effects, we
dwell now on the uncertainty of predicting the rate in the
kinematics where $b\ra u$ decays can be experimentally disentangled
from
the KM-allowed transitions. In other words, we must understand how
accurately
one can calculate theoretically $\Phi(M_{\rm max})$ for realistic
values
of the cutoff mass $M_{\rm max}$. The first step is clearly to
analyze
its dependence on not yet precisely known parameters of the
heavy quark expansion.

In Fig.~4 the predictions are plotted as a function
of $M_{\rm max}$ for the three values
$\tilde \mu _{\pi}^2= 0.2, \; 0.4, \; 0.6\GeV^2$ with
$\tilde m_b = 4.82\GeV$ fixed and in
Fig.~6 for $\tilde m_b = 4.78\GeV,\;4.82\GeV$ and $4.87\GeV$ with
$\tilde \mu _{\pi}^2= 0.4\GeV^2$ fixed. We see immediately that
the lower $M_{\rm max}$ is chosen, the {\em higher} the sensitivity
to long-distance dynamics becomes. This is not surprising
qualitatively since for a lower $M_{\rm max}$ a smaller fraction of
the overall rate is included.

The intervals of variation for $m_b$ and $\mu_\pi^2$ we consider are
quite conservative. The actual existing uncertainty in the latter
seems
to be $1.5$--$2$ times smaller \cite{rev};
it is hardly possible that the $b$ quark
mass entering our calculations can vary in a wider interval. Taking 
for orientation
$M_{\rm max} = 1.5$ to $1.6\GeV$ we observe that the dependence
of
the fraction of the events $\Phi\left(M_{\rm max}\right)$ on $m_b$
is
more pronounced. To illustrate this dependence more clearly, we
give the exploded version of Figs.~5 and 6 in Figs.~7 and 8 where
we vary $m_b$ by twice larger amount $\pm 100\MeV$. The variation in
the
fraction of the events remains reasonably small for $M_{\rm
max}\gsim 1.6\GeV$.

The stability deteriorates when one descends already a few hundred
$\MeV$
below. 
One not only loses statistics, the fraction $\Phi$ becomes
essentially
dependent on the nonperturbative parameters.

Thus we see that the goals of larger statistics and smaller
sensitivity to nonperturbative effects favor the selection of as
{\em high} a value for $M_{\rm max}$ as experimentally feasible,
considering leakage from $b\ra c$.

To be more specific: taking the plots literally, about $70\%$
of all $b\ra u$ events lie
below $M_{\rm max} =1.5\GeV$ with a spread of $\pm 8$\%; for
$M_{\rm max} =1.3\GeV$ and $1.0 \GeV$ these numbers read
$\simeq 60\%$ and $\simeq 35\%$, respectively, with a
spread of $ \pm 12\%$ and $ \pm 15\%$,
respectively. These are uncertainties on top of those
for the total $B\ra l \nu X_{u}$ width listed in
Eq.~(\ref{VUBTOTAL}) which are encountered in converting the
semileptonic width into $|V_{ub}|$.
For $M_{\rm max} \lsim 1.5\GeV$ the uncertainties
in the width pale in comparison to those
due to the shape of spectrum.

In reality, one must allow for additional uncertainties associated
with
other approximations made in the analysis; most profoundly, the
effect
of the subleading nonperturbative effects in the structure functions
not
captured by naive Fermi motion is quite non-negligible for the actual
$b$
quark mass. Their theoretical discussion and the estimate of related
uncertainty can be found in \cite{bsg,dist}. As long as 
one does not aim for an {\em
absolute} precision in $\Phi(M_{\rm max})$ below $\sim 10\%$, 
they are not expected to be important. Nonetheless, 
additional work toward better control of remaining uncertainties is
clearly needed. Without a doubt, it will receive necessary
theoretical
attention as soon as the experimental feasibility of such
measurements
will be established and their ultimate precision is understood.

\section{Summary and Outlook}

We have shown here that if one succeeds in measuring the
hadronic recoil mass spectrum in semileptonic $B$ decays, one
can extract $|V_{ub}|$ in a theoretically clean and
accurate way. In our theoretical analysis we split the
problem into two parts -- one is calculating the overall
semileptonic
$b\ra u$ width of $B$ mesons, and the second is determining the
fraction of $b \ra u$ events that can be discriminated kinematically 
against $b\ra c$ processes.
A
good theoretical control over the first theoretical ingredient
allowed
us to concentrate on the more subtle second aspect.

Perturbative as well as nonperturbative QCD
dynamics have a large impact on the mass spectrum -- they
broaden it considerably. Yet the numerical analysis suggests that
they do not change very much one basic
feature, namely that only a small tail distribution extends
beyond $M_D$. Furthermore, while the shape of the
spectrum is sensitive to long-distance dynamics, its integrated
weight,
say below $1.5\GeV$ is much less so. There is nothing magic about
the
value of $1.5\GeV$. 
What is important  is that one selects a cut-off
that is
high enough for most of the $b\ra u$ transitions to contribute below
and low enough for almost all $b\ra c$ leakage to occur above.
Nevertheless, not only statistics, but also the issue of theoretical
reliability calls for attempts to set as high a cut-off as possible.

Thus the following scenario emerges: measuring the semileptonic
decay rate
below, say, $M_{\rm max}=1.5 \GeV$ will enable us to extract
$|V_{ub}|$
with a theoretical uncertainty that does not exceed $10\:$--$\:20$
percent. An overall uncertainty of below $10 \%$
appears achievable or at least not impossible since with
$10^7 - 10^8$ $B$ decays and
${\rm BR}(B \ra l \nu X_u)\simeq 0.0015$ one estimates
that a {\em statistical} accuracy level $\sim {\cal O}(1\%)$
would be available, provided a large fraction of the decay events can be
successfully utilized. As described above, a particular premium
above and beyond statistics has to be placed on achieving a high
value for $M_{\rm max}$, not below $1.5\GeV$.

Meanwhile, if the studies will demonstrate the feasibility of such
experimental measurements, the theoretical description can be improved.
The feasible method of obtaining accurate perturbative corrections to
the decay structure functions was suggested in \cite{bsg}, the so-called
APS which combined advantages of exact one loop-expressions,
exponentiation of soft infrared/collinear effects and incorporating of
the running of the coupling. It intrinsically includes the normalization
point and thus is free of problems associated with
using the pole mass of $b$ quark encountered in the ``practical'' OPE.
Yet for the reasons elucidated above we do not anticipate qualitative
changes in our predictions. In particular, the resummation of
next-to-leading logs does not seem very promising in the semileptonic
decays due to a limited gap in the momentum scales. Without a cutoff,
however, the resummation of leading logs seems technically necessary not
to violate unitarity for soft emissions. It is important to check that the  
sensitivity to the exact treatment of the perturbative corrections
is not high, since application of the perturbative expansion at a
precise level for the soft gluons with $M_x \lsim 1\GeV$ seems
problematic.

In the near future one will presumably determine
$m_b$ and $\mu _{\pi}^2$ from $b\ra c$ transitions with even better
accuracy than stated above. Lastly, with increasing statistics
one can measure the $b\ra u$ recoil mass spectrum with good accuracy,
rather than its integral as a function of $M_{\rm max}$. This in
turn
will allow us to independently place bounds on $m_b$ and $\mu
_{\pi}^2$
in  a systematic way. Comparing these bounds with other
determinations of
these basic quantities will serve as an essential quality check on
systematic uncertainties and the quantitative applicability of the heavy
quark expansion.

If accurate measurements of the recoil mass spectrum are feasible at all, it 
is advantageous to perform such studies separately for charged and
neutral $B$. In the KM-suppressed semileptonic decays one can observe
the effects of Weak Annihilation as the difference in the decay
characteristics for charged and neutral $B$. Moreover, in the decays to
massless leptons one is sensitive to the nonfactorizable parts of the
expectation values of the four-quark operators responsible for the
effects of Weak Annihilation \cite{WA}. On the other hand, it is just
the theoretical expectations for nonfactorizable parts which are a subject 
of some controversy in the literature at the moment. Gaining direct
experimental information not only can put theoretical predictions of 
beauty lifetimes on more solid grounds, but
also has an independent theoretical interest. Ultimately, Weak
Annihilation can even introduce an element of theoretical uncertainty in
the extraction of $|V_{ub}|$ in the discussed way unless its effect is
controlled through such studies.

At the same time, theoretically, the effects of Weak Annihilation are 
expected to come from small $M_X$ (or even from small hadronic
recoil energy $E_h$) \cite{WA}; 
singling out this domain can amplify the effect
and make it detectable even if its overall scale is 
insignificant. \vspace*{.3cm}\\
{\bf Acknowledgments:} \hspace{.4em} After publication of paper
\cite{dist} where the QCD-based 
analysis of the recoil mass distribution was first 
reported, along with other decay distributions, a paper by Falk, Ligeti and 
Wise appeared \cite{flw} which discussed 
the similar issue of extracting $|V_{ub}|$ from the recoil mass spectrum. 
R.~Dikeman would like to thank B.~Urbanski for her generous 
hospitality during the completion of this work.
Our work was supported in part by
the National Science Foundation under the grant number PHY 92-13313
and
by DOE under the grant number DE-FG02-94ER40823.

\newpage

\begin{figure}
\vspace{8cm}
\includegraphics{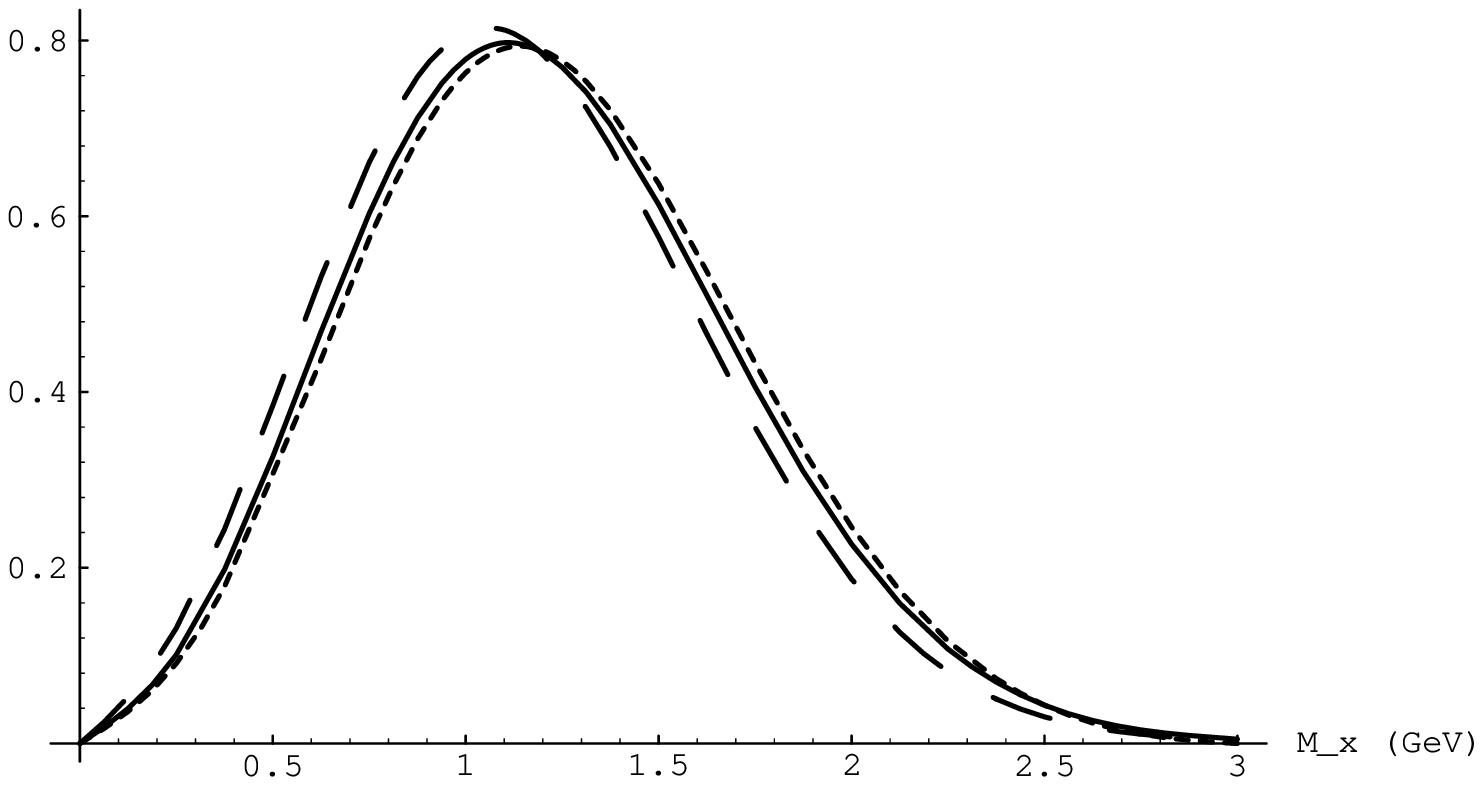}
\caption{
Effect of radiative corrections on $d\Gamma/dM_x$:
solid line shows $\as=0.3$, dashed 
and dotted lines are $\as=0.1$ and
$\as=0.5\,$, respectively.
$\tilde m_b=4.82\GeV$ and $\tilde \mu_\pi^2 = 0.4\GeV^2$.
}
\end{figure}

\begin{figure}
\vspace{8cm}
\includegraphics{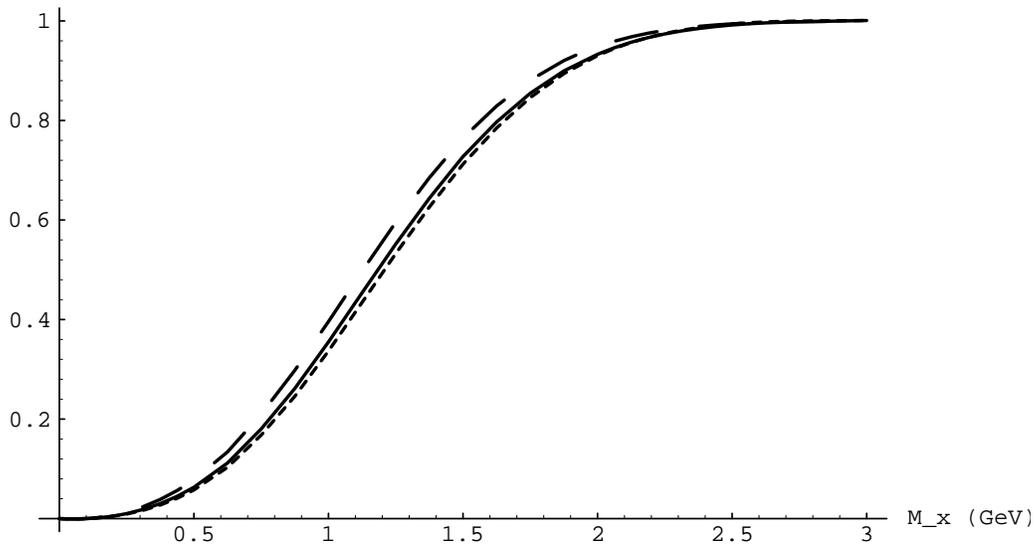}
\caption{
The integrated fraction of the events $\Phi(M_x)$. 
}
\end{figure}

\newpage

\begin{figure}
\vspace{8cm}
\includegraphics{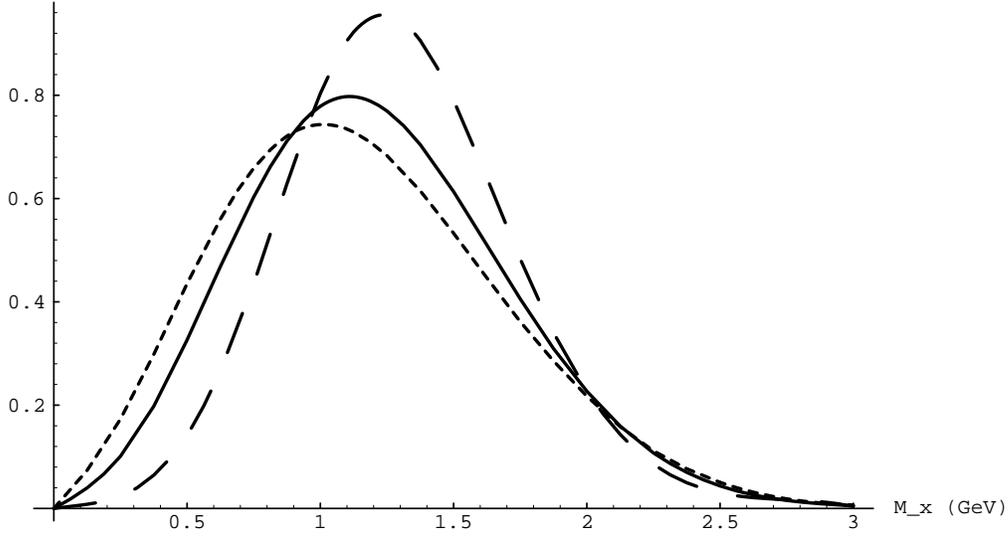}
\caption{
Dependence of $d\Gamma/dM_x$ on $\mu_\pi^2$:
long-dashed, solid,
and short-dashed lines correspond to $\tilde \mu_\pi^2 =0.2\GeV^2$, 
$0.4\GeV^2$
and $0.6\GeV^2$. The $b$-quark mass $\tilde m_b=4.82\GeV$, $\as=0.3$.
All
distributions are normalized to the same total width.
}
\end{figure}

\begin{figure}
\vspace{8cm}
\includegraphics{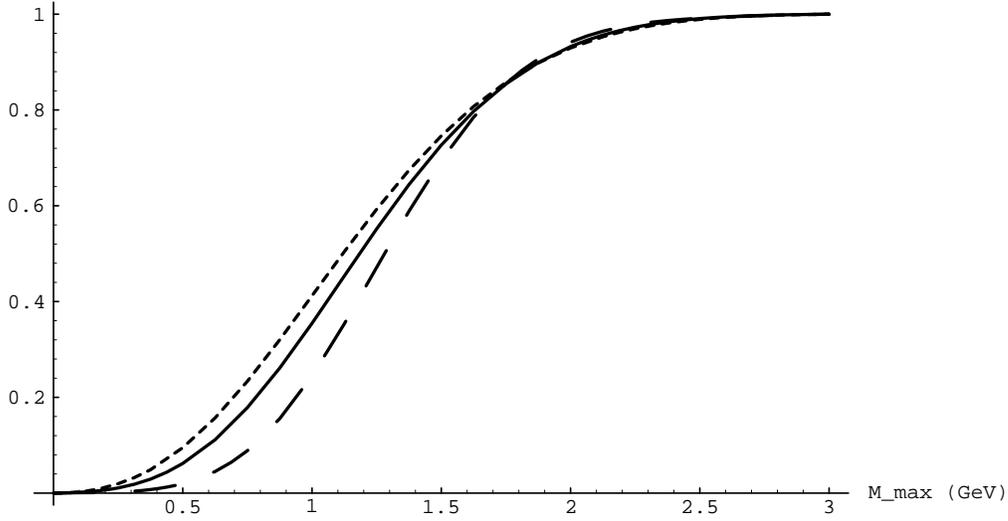}
\caption{
$\Phi(M_x)$ for $\tilde \mu_\pi^2 =0.2\GeV^2$, $0.4\GeV^2$
and $0.6\GeV^2$.
}
\end{figure}

\newpage

\begin{figure}
\vspace{8cm}
\includegraphics{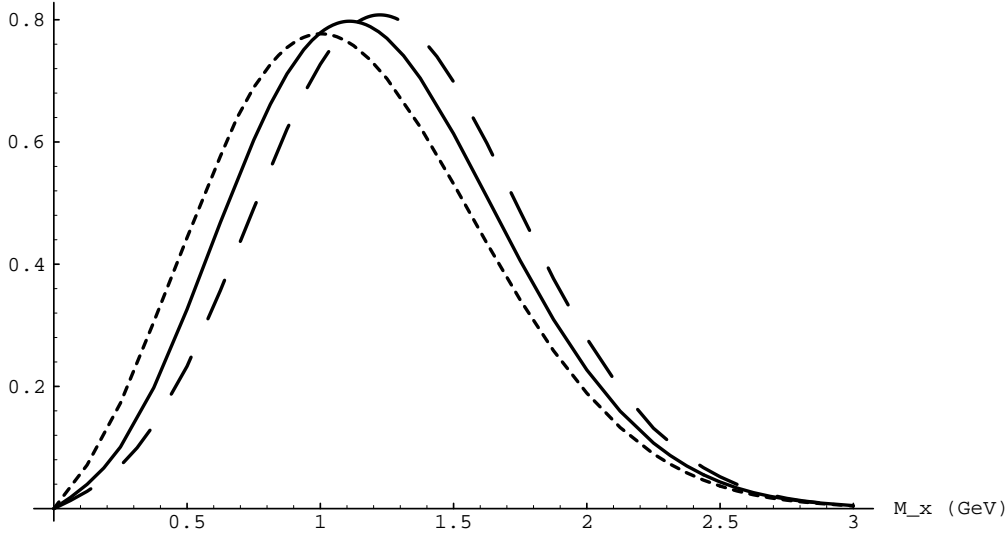}
\caption{
Dependence of $d\Gamma/dM_x$ on $m_b$:
long-dashed, solid,
and short-dashed lines are $\tilde m_b =4.77\GeV$, $4.82\GeV$
and $4.87\GeV$. $\tilde \mu_\pi^2 =0.4\GeV^2$, $\as=0.3$.
}
\end{figure}

\begin{figure}
\vspace{8cm}
\includegraphics{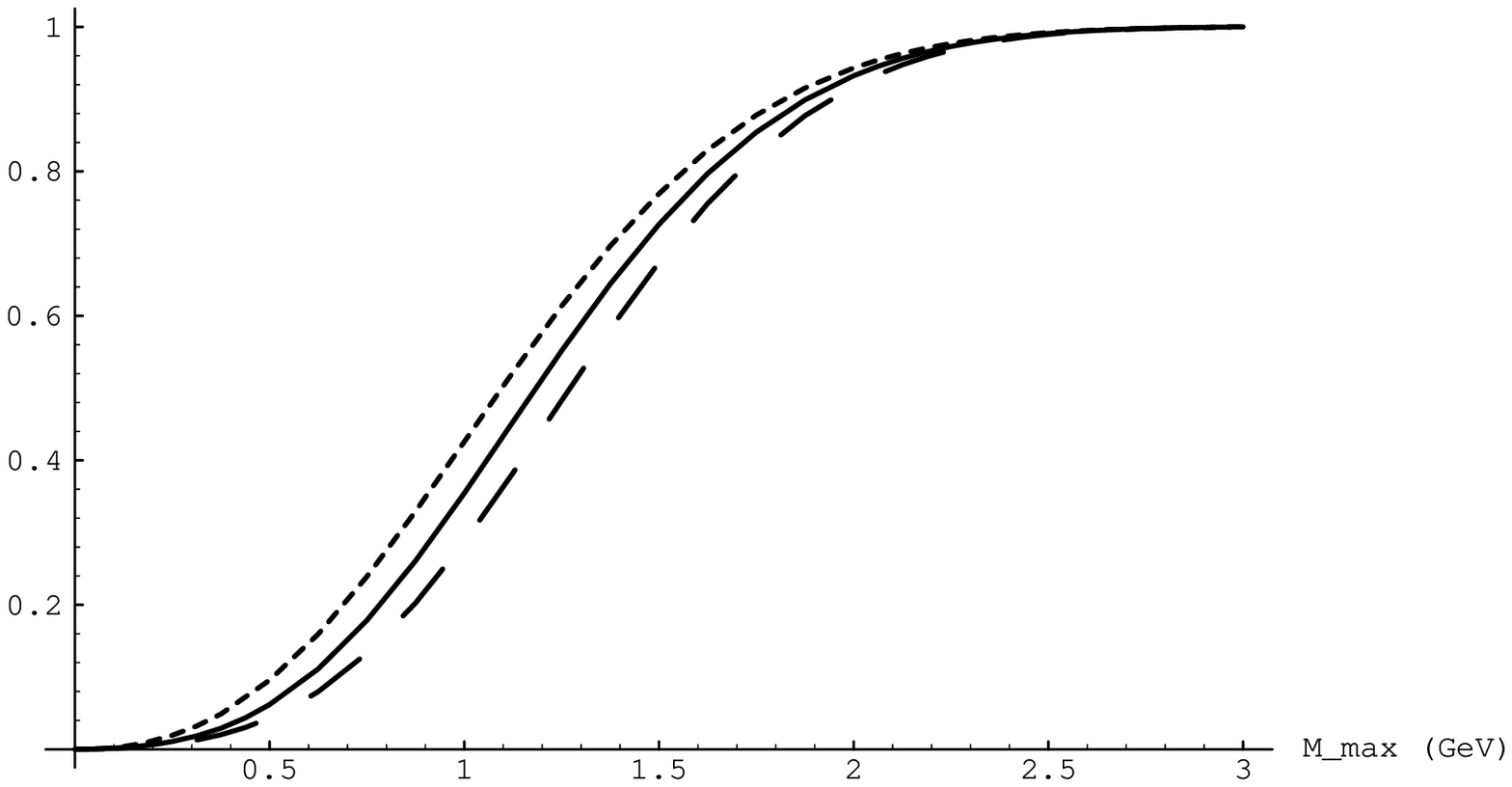}
\caption{
$\Phi(M_x)$ in the same setting as in Fig.~5. 
}
\end{figure}

\newpage

\begin{figure}
\vspace{8cm}
\includegraphics{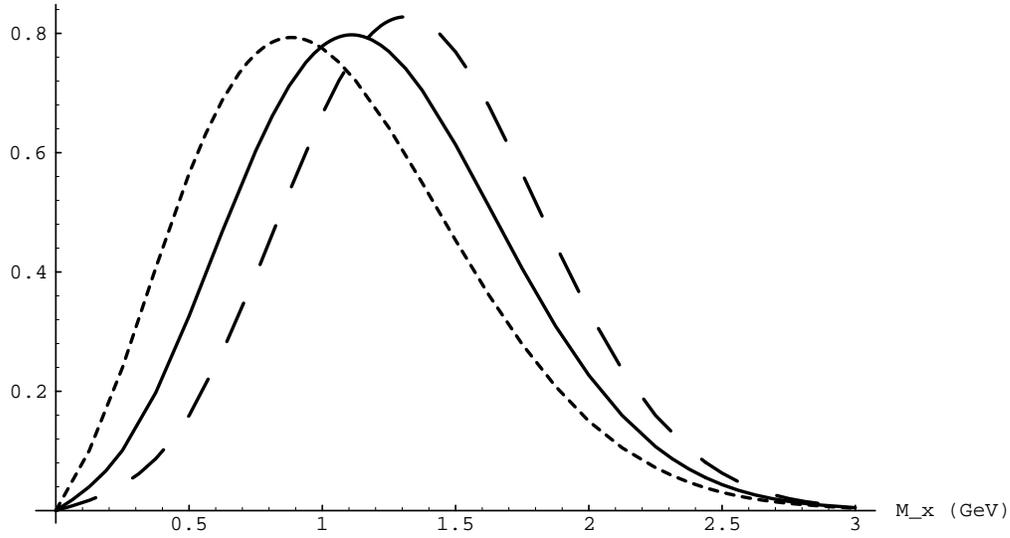}
\caption{
$d\Gamma/dM_x$ for 
$\tilde m_b =4.72\GeV$, $4.82\GeV$
and $4.92\GeV$. $\tilde \mu_\pi^2 =0.4\GeV^2$, $\as=0.3$.
}
\end{figure}

\begin{figure}
\vspace{8cm}
\includegraphics{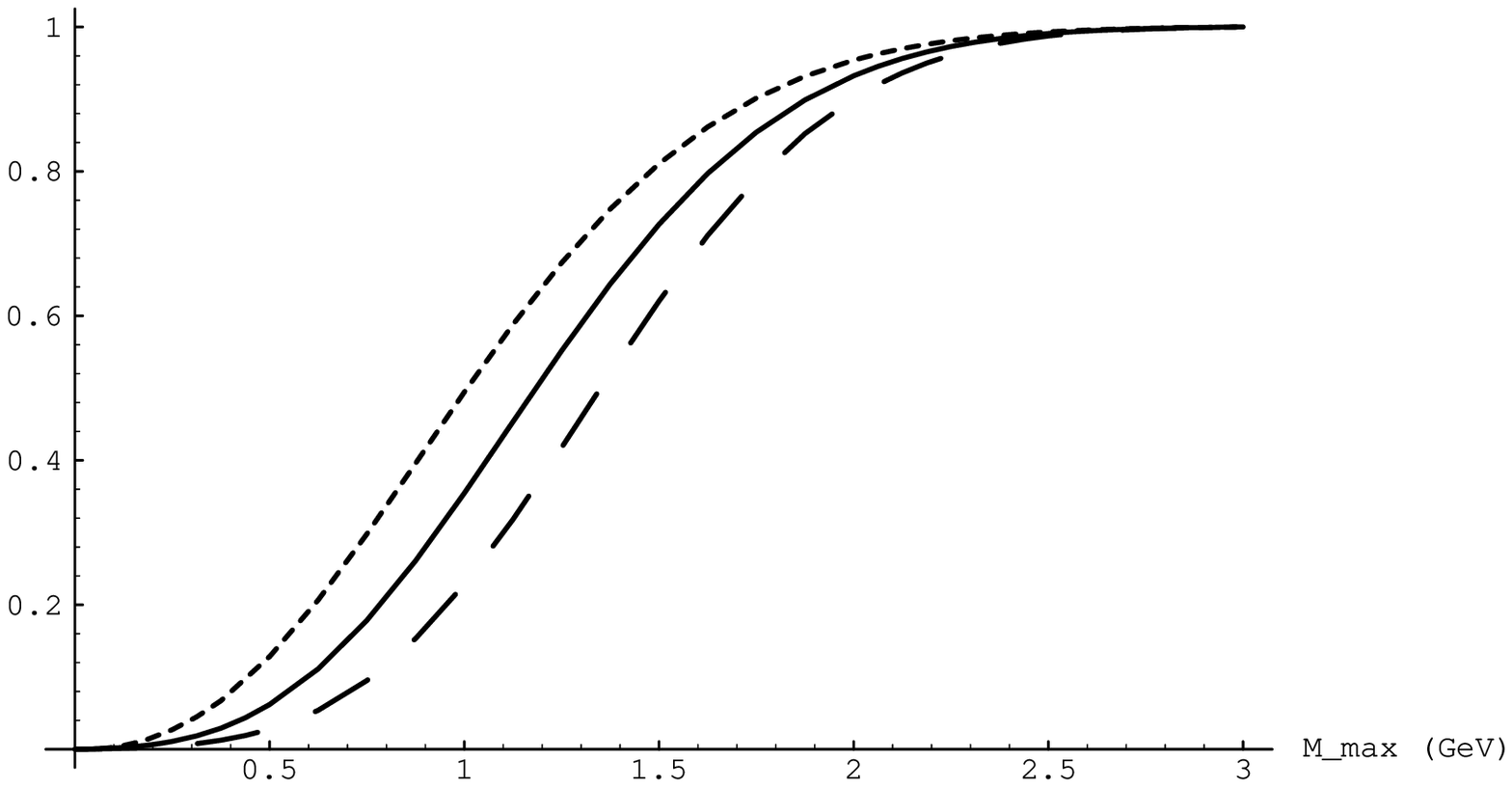}
\caption{
$\Phi(M_x)$ in the same setting as in Fig.~7.
}
\end{figure}


\begin{thebibliography}{99}
\bibitem{tools}
For a recent review see \cite{rev}; the details of the QCD descriptions of 
the  ``Fermi motion'' in beauty decays can be found in the dedicated 
papers \cite{motion,bsg} and other publications \cite{rand} referred to there.
\end{thebibliography}
\end{document}